# Scalability, Availability, Reproducibility and Extensibility in Islamic Database Systems

Umar Siddiqui [1], Habiba Youssef [2], Adel Sabour [3], Mohamed Ali [4]

[1,2,3,4] School of Engineering and Technology, University of Washington, Tacoma, USA
[1] umarqr@uw.edu, [2] hyoussef @uw.edu, [3] sabour@uw.edu, [4] mhali@uw.edu

## Abstract
With the widespread of software systems and applications that serve the Islamic knowledge domain, several concerns arise. Authenticity and accuracy of the databases that back up these systems are questionable. With the excitement that some software developers and amateur researchers may have, false statements and incorrect claims may be made around numerical signs or miracles in the Quran. Reproducibility of these claims may not be addressed by the people making such claims. Moreover, with the increase in the number of users, scalability and availability of these systems become a concern. In addition to all these concerns, extensibility is also another major issue. Properly designed systems can be extensible, reusable and built on top of one another, instead of each system being built from scratch every time a new framework is developed. In this paper, we introduce the QuranResearch.Org system and its vision for scalability, availability, reproducibility and extensibility to serve Islamic database systems.

## 1. Introduction
How many times do we see developers who are enthusiastic to build applications to serve the holy Quran, yet, they do not pay attention to the accuracy of the Quran text they present? How many times do we see people spending hours of data entry to build a database of the Quran text, yet, this database remains a local copy on the author's personal computer? How many times such databases are eventually lost or at least do not get enough publicity? How many times do we see individuals claiming scientific and numerical miracles of the Quran, yet, we discover over time that these claims are fallacious, erroneous or distorted? How many times do we see applications producing statistics related to the Quran text, yet, we do not see much effort invested in other computing fields? Would not it be the time to explore how natural language processing, data mining, machine learning and information retrieval (to name a few) can serve the holy Quran? We can continue to list more challenges that are interfering with the advancement of the state of the art in the computational analysis of the Quran text.

Inspired by a previous system the authors built in the domain of environmental sciences, called AMADEUS (Hendawi et al, 2014; AMADEUS, 2022), the authors extended the idea to the direction of Islamic Databases. AMADEUS stands for Azure Marketplace of Applications for Diverse Environmental Use as a Service. This system has the ability to consume data from a wide variety of environmental data sources, and to store them in one integrated data store. To guarantee the availability and accessibility of this data, AMADEUS hosts all data in the cloud. The system gives the users the luxury to query and analyze the underlying data sets by writing SQL queries and/or through the system's user-friendly graphical user interface. The system also gives the users the ability to study the correlation between multiple data sets using the algorithms and data structures natively offered within the framework. The users of the system share and exchange not only datasets but also libraries of code and modules of various machine learning algorithms.

QuranResearch.ORG (QuranResearch, 2022) aims to provide the space and tools that ease, support and integrate the efforts of researchers in the field of Quran computational research. QuranResearch.ORG provides a public backend database server that contains an indexed representation of the Quran text in addition to a set of general-purpose tools and packages to perform counting, searching and mining of the Quran text. QuranResearch.ORG invites Researchers to issue their queries against our public backend database server and to, then, retrieve/analyze/share their results through the system's frontend. The database schema is documented and published publicly on the system's website (QuranResearch Documentation, 2022). The proposed system is designed to serve as a "wiki" style repository for all the results that researchers would obtain and reflect on from their efforts in Quranic research. We hope to provide a framework for Quranic research in the fields of Textual Analysis, Numeric Analysis, Linguistic Analysis and Natural Language Processing (NLP).

QuranResearch.ORG expects two types of audience. QuranResearch.ORG hopes to inspire researchers with software development skills to conduct research in the field of Quran textual, numerical and linguistic analysis. Then, researchers are expected to publically share their results seeking comments, feedback and suggestions from the general public. While researchers would constitute the first group of audience, the general public would constitute the second group of audience who would be interested in the messages, observations and miracles in the book of Allah, the Quran.

The rest of the paper is organized as follows. Section 2 overviews the features of the proposed system. Section 3 presents the system's user interface that targets the general public and explains how it provides a portal for all the queries that are published by the contributing researchers/developers. Section 4 discusses the researchers/developers dashboard and how a developer would author a query and, then, request the query to be published so that it becomes publicly available to the world. The paper is concluded in Section 5.

## 2. QuranResearch.ORG Features

QuranResearch.org is designed to achieve several goals. In this section, we summarize the system's main features that help achieve these goals. Please note that QuranResearch.org is an ongoing project and its development team is working to add and extend these goals over time. This paper provides a high level description of the system's features and vision without diving deep into the technical details of how the system achieves these goals.

### 2.1 Accessibility

The system provides a database of the Quran text that is publicly available and open to all researchers. Researchers issue their queries through the QuranResearch.ORG website and count on the backend server to run their "possibly" long-running queries, then notify the users once the results are ready. There is no need for hardware or software installation on the researcher's side.

### 2.2 Efficiency and performance

The system provides a highly indexed representation of the Quran text that speeds up the search and analysis of the Quran text. The system aims at scaling up to multiple concurrent users performing long-running queries on the Quran text. Elasticity in computation by hosting the system in the cloud enables adding servers to support query workloads at peak times.

### 2.3 Accuracy

The system aims to provide an accurate database for the whole Quran, with accurate information and statistics. The system aims to provide a database that grows over time to include multiple Quran narrations and different counting or numbering methods of Ayahs. The system aims to support true claims around "numeric miracles in Quran" with concrete numbers and, meanwhile, refute false claims that are imprecise and untruthful. Authentic and accurate databases come at the heart of the system to achieve this goal.

### 2.4 Application Programming Interfaces (APIs), tools and packages

The system provides an open platform for researchers to issue their queries against the backend database and retrieve results instantaneously or be notified once results are ready. The system is designed for developers who are comfortable with their programming skills and who are willing to use these skills in Quranic research. The system provides Application Programming Interfaces (APIs) with various settings and tuning knobs to easily search and retrieve words/ayahs. The system provides layers of data analysis, data mining and data visualization on top of the Quran database. QuranResearch.Org does not plan to be only accessible through its UI website but to be also accessible through its APIs that software developers can connect to and issue their queries against the system's backend.

### 2.5 Knowledge sharing and research network

QuranResearch.org is expected to be the "Wikipedia-like hub" that lists all the results from the collected efforts of the community of Quran researchers in one central place. QuranResearch.org is expected to build a network of researchers in the field, where they can communicate, provide feedback and collaborate in research. The website does not only aim to be a hub for sharing results but also to be a hub for researchers to share tools, pluggable modules and packages that researchers can use and integrate in their research. Consequently, the "value of the whole system becomes greater than the sum of the parts".

## 3. The System's Front End

This section explains how to navigate through the QuranResearch.org frontend, describes the functionality provided by each page and shows how the backend features are surfaced to the end user.

### 3.1. The System's UI structure and the Basic Navigational Functionalities

As shown in Figure 1, the user interface is divided horizontally into three sections. The leftmost section and the top part of the middle section provide the "basic navigational controls". It allows the user to navigate to the desired place in the Quran by surah number, chapter (juz') number, page number, etc. Moreover, some additional navigational controls allow the user to navigate forward and backward by one, ten or hundred pages with every click.

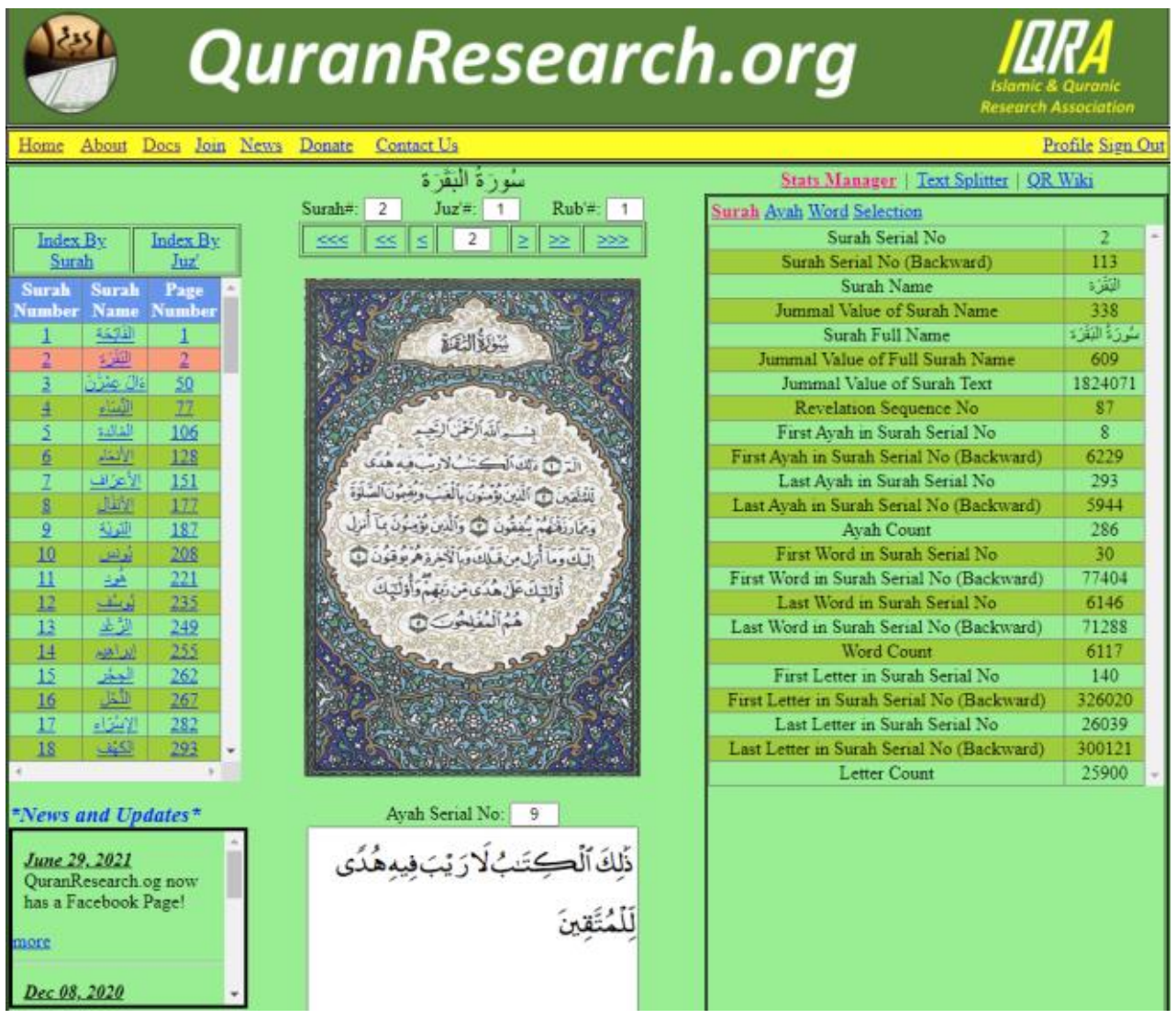

*Figure 1. Homepage of QuranResearch.org.*

The middle section displays the Quran page that is navigated to by the user. Clicking an ayah in the displayed page will populate a textbox at the bottom of the middle section with the text and the ayah number of the ayah that has been clicked. This textbox (at the bottom of the middle section) is referred to as "Selected Ayah Textbox". The middle section is basically the Quran page and, more specifically, the selected ayah that subsequent queries will be considering. Note that the Quran page and the selected ayah (displayed in the "Selected Ayah Textbox") are navigated to through the leftmost section, the navigational buttons at the top of the middle section and by clicking on ayah in the displayed Quran page. Moreover, the Quran page and the selected ayah can also display (as explained in the next paragraph) the page and ayah linked to a row in the query result of the query being executed in the rightmost section of the page.

The rightmost section of the user interface is dedicated to the operations and queries being executed over the Quran text. All text operations, numerical operations, analysis queries and visualizations are displayed in the rightmost section of the interface. These operations are expected to be executed over the selected page, ayah, part of an ayah, word, or even part of an ayah that is highlighted in the "Selected Ayah Textbox". The rightmost section is featured by the Statistical Manager, the Text Splitter and the Quran Wiki modules of the system. These modules are described in the following section.

### 3.2. Stats Manager, Text Splitting and Quran Wiki

In the rightmost section of the UI, comes the statistical and textual analysis components of the Quran. There are three tabs in the interface, namely:

**Stats Manager**: As shown in Figure 2, this subpage displays statistics related to the current surah, ayah, word, or selected text of the Quran. It gives the serial numbers and counts of Ayahs/Words/Letters, counting forward/backward (from beginning or end) relative to the beginning/end of the whole Quran, current Surah/Ayah/Word. The statistics can be computed according to four granularities: a full surah, an ayah, a word, or a selection (which is the text highlighted by the user in the Selected Ayah Box).

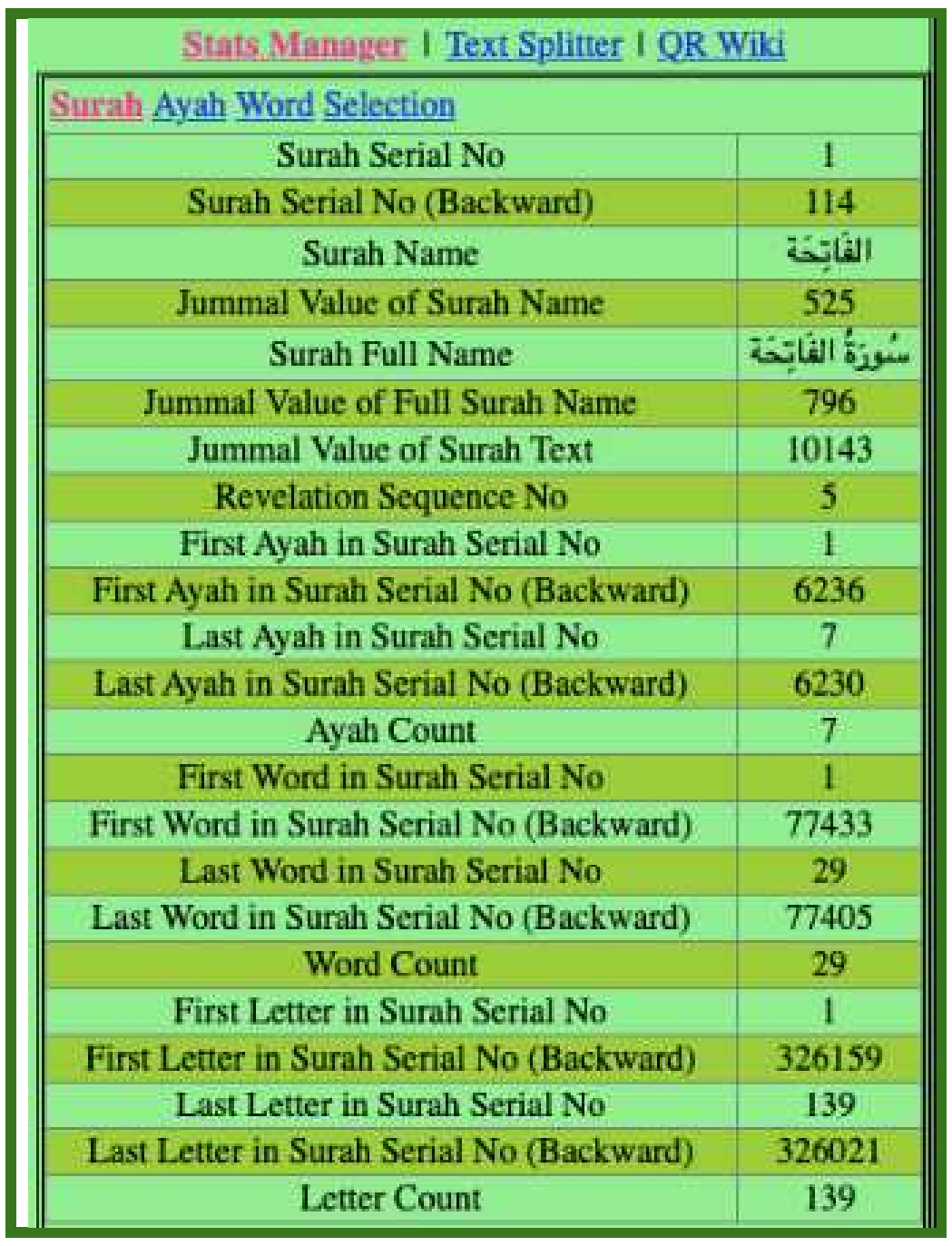

*Figure 2. The Statistics Manager.*

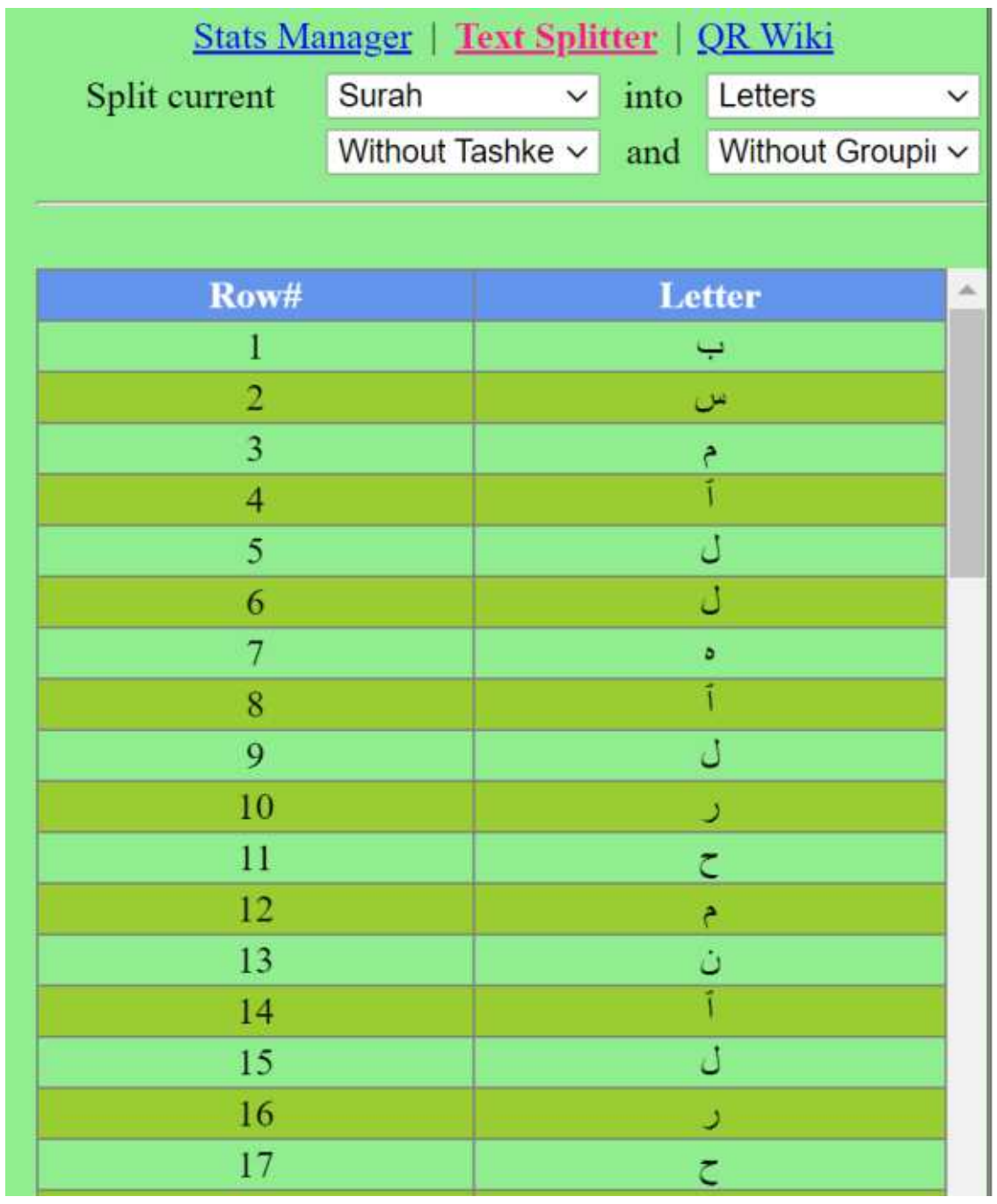

*Figure 3. The Text Splitter.*

**Text Splitter**: The text splitter (Figure 3) splits the current surah, ayah or word into its constituting letters or words. The user has the option to either consider Tashkeel or not while splitting. The user can also group similar words/letters and get a count of how many times the word/letter is repeated in the currently selected Surah/Ayah. The text splitter helps identify how the words and letters are counted by various queries through showing the constituting parts of each ayah or word.

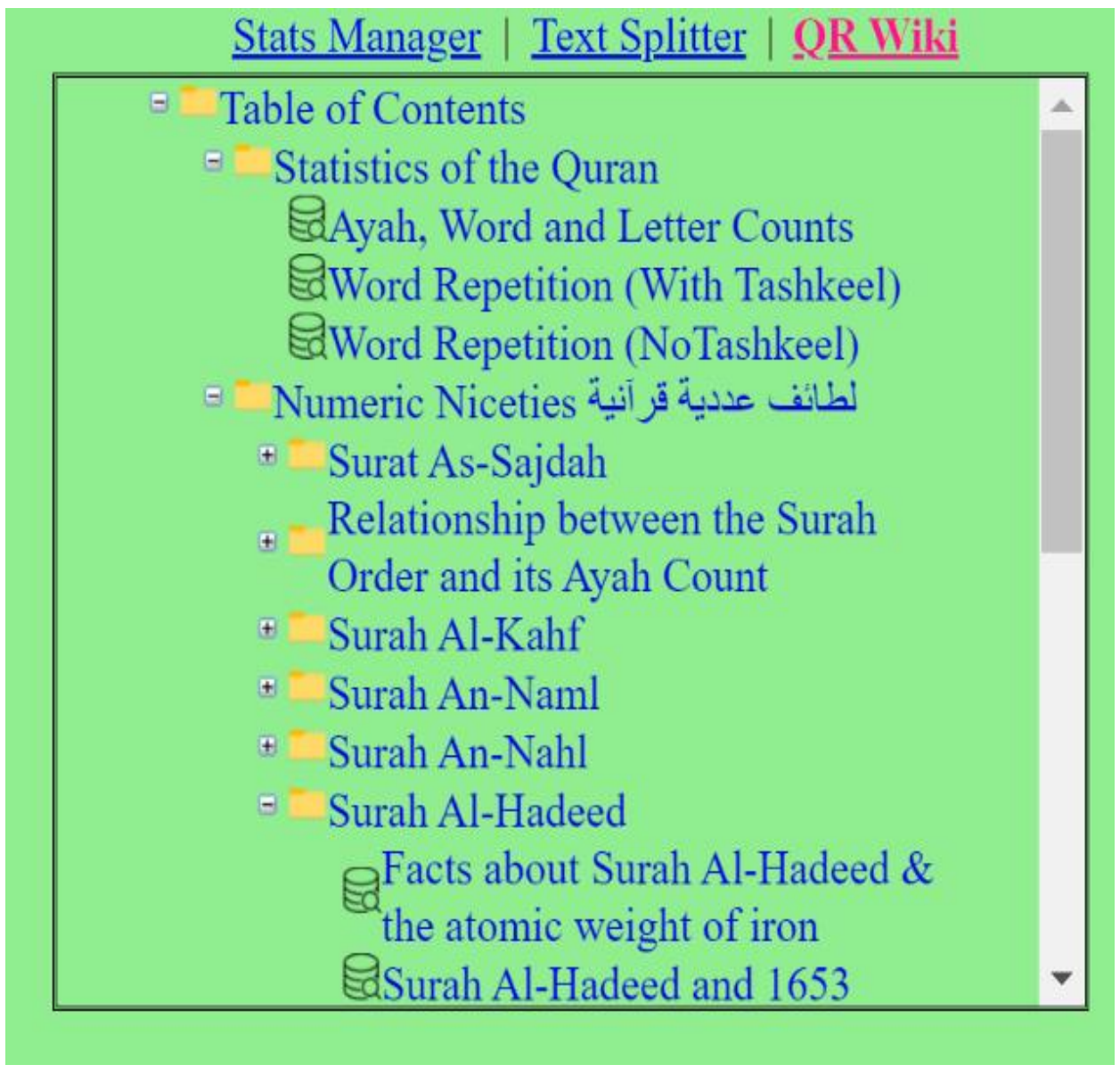

*(a) Table of Contents*

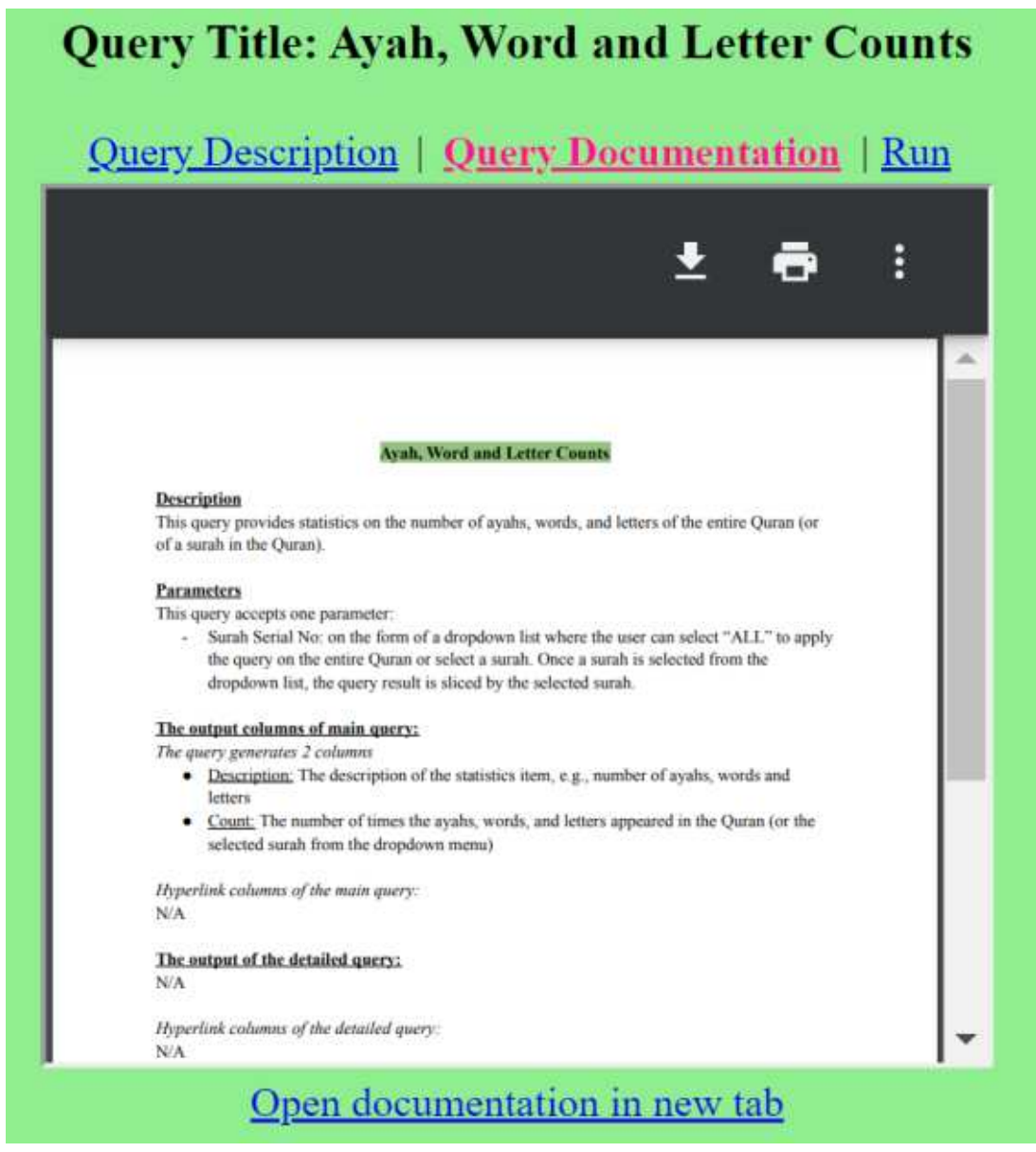

*(b) The Query Documentation*

*Figure 4. The QR Wiki.*

**QR Wiki**: This part is definitely the most important section of the system. All applications that provide statistical information of the Quran text are "static". This component of the system is what makes the system dynamic and extensible. Researchers and developers from the community can write their Queries in the SQL Language through the developer's corner (as described in Section 4). Then, the developer has the option to request the publication of the query on the website, after adding the proper documentation. All published queries will appear under the QR Wiki section, which lists and classifies all developer-written queries that have been approved for publications by the system administrators.
QR Wiki is expected to be the "Wikipedia-like hub" that lists all the results from the collected efforts of the community of Quran researchers in one central place. As shown in Figure 4, QR Wiki provides a table of content (Figure 4(a)) that classifies queries into a tree structure based on topics and sub-topics. Once a query is chosen, the query description, the query documentation, the query's source code in SQL and the query result are displayed through various tabs available at the top part of the query interface (Figure 4(b)). As the Quran research community grows bigger, the QR Wiki is expected to grow larger and to accommodate more valuable queries. QR Wiki should facilitate the discussion among the research community around the published queries. Moreover, QR Wiki is expected to create a healthy environment for sharing ideas, reviewing results and suggesting improvements.

## 4. The Developer's Dashboard

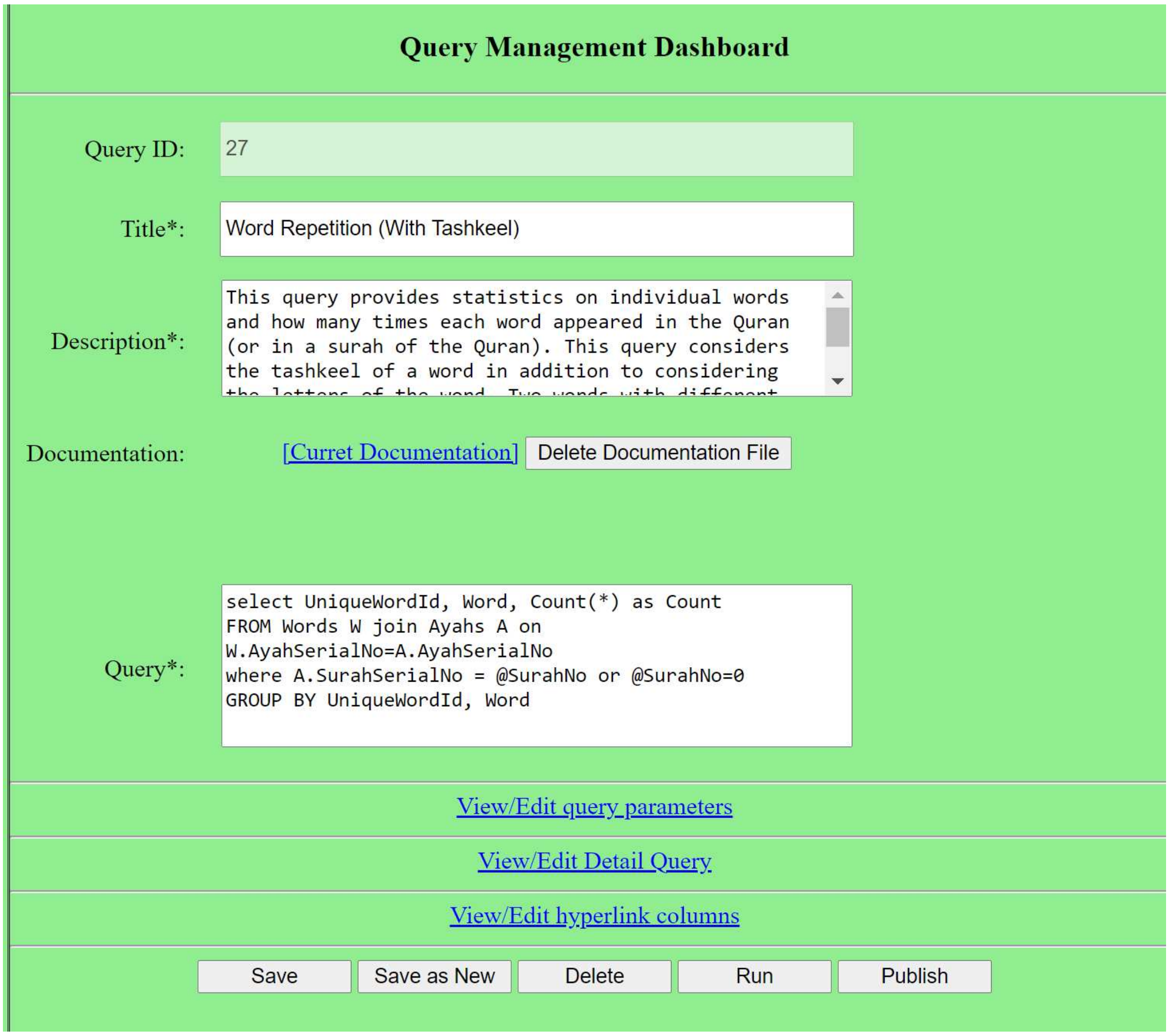

*(a) The query title, description and main SQL query.*

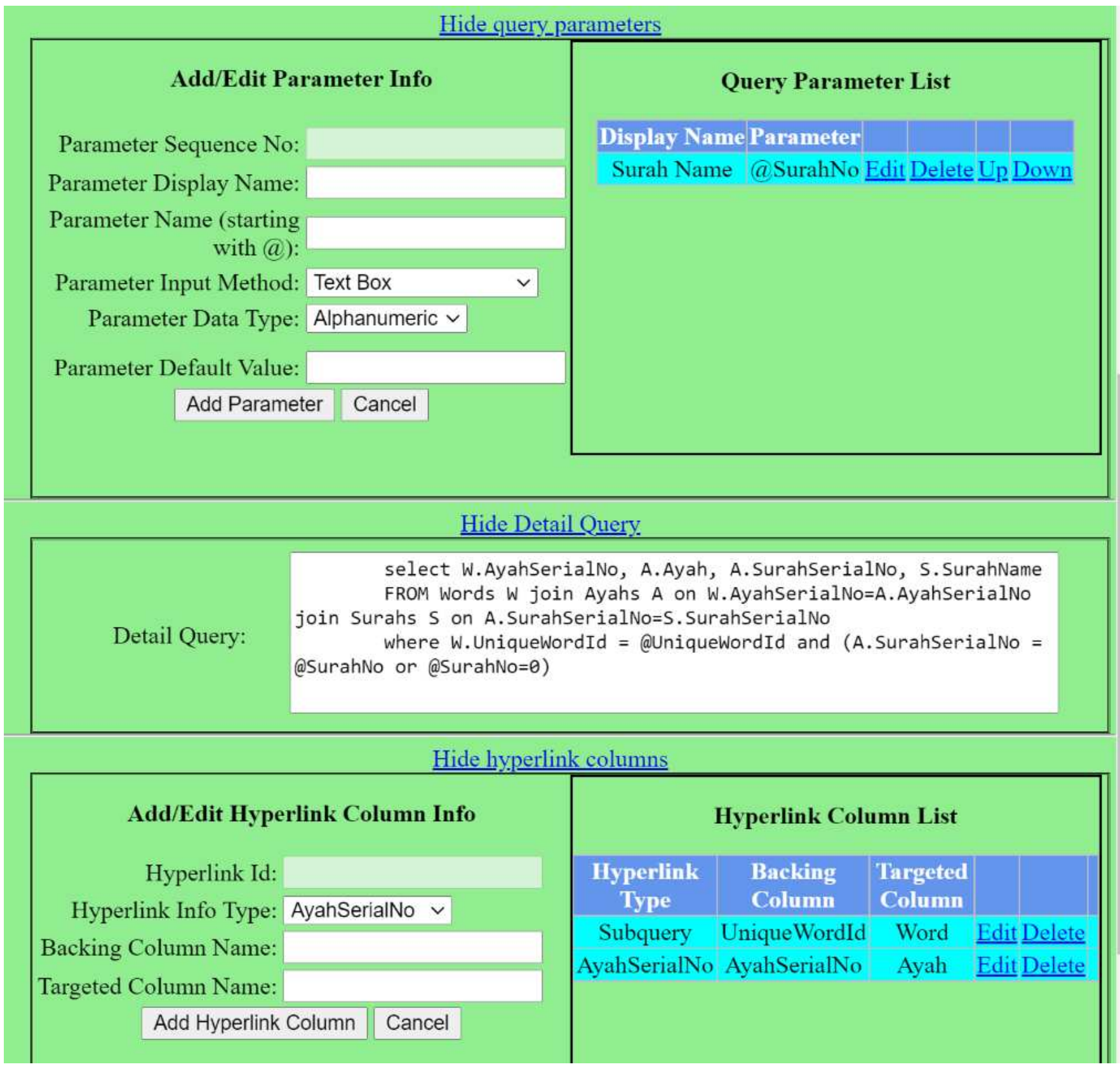

*(b) The query parameters, detail query SQL and hyperlink columns.*

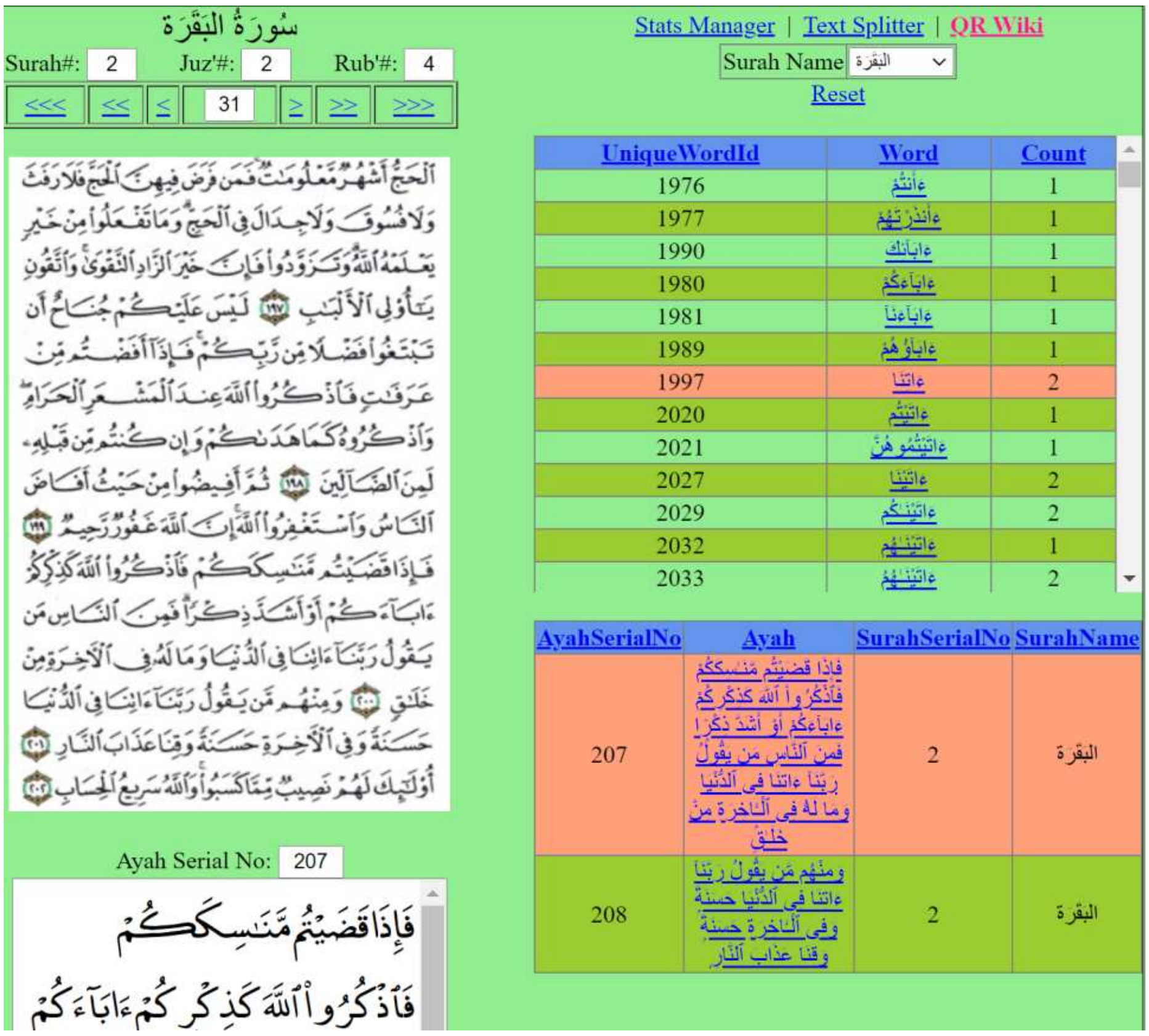

(c) An example query result.

*Figure 5. The Developer's dashboard.*

As mentioned earlier in the paper, the proposed system has two types of audience: (a) The developers who have the programming skills and can write queries to extract insights and statistics from the Quran text and (b) the general public who would consume the result of the queries. The developer's side of the system is crucial to the extensibility of the system. Unlike existing applications that provide statistics on the Quran, QuranResearch.Org provides a platform for developers all over the world to envision their queries and to inspire their creativity about what type of insights to extract from the Quran text. This section provides the interface of the developer's dashboard and the culture behind it.

The developer will author a query in a system-accepted language. As of now, the system accepts SQL as its language. However, future plans include the possibility of accepting multiple languages to query the Quran database, with a possibility of a graphical editor to compose queries. To write a query, the user would provide (through the developer's dashboard as in Figure 5) the following information:

- Title, which is a descriptive identification of the query.
- Description, which documents the query in a paragraph or two to highlight the main idea of the query.
- Documentation, where the user uploads a PDF file that documents the query details.
- Query (or main Query), which is an SQL query that generates the query results.
- Parameters, where the query can accept various sorts of parameters that slice the query result, for example, by surah, chapter (juz') or ayah.
- Detail Query, which is another query that is executed once a row from the Main Query result is clicked. The Detail Query is linked to the Main Query through a hyperlink column. Once the hyperlink column in a row (in the Main Query result) is clicked, the value of that cell is passed to the *Detail Query* as a parameter. The result of the Detail Query is displayed under the main query result.
- Hyperlink columns, which are used to provide hyperlinks in the query results. These hyperlink columns serve as a linking mechanism between the Main Query result and the Detail Query. Also, the hyperlink columns serve as links in the query result so that the user can transition to the page and/or ayah that is referenced in the query result.

Figure 5a shows the developer's dashboard of a typical query with the title, description, documentation and Main Query sections while Figure 5b expands the query parameters, the Detail Query and the hyperlink columns sections. The QuranResearch.Org website provides a series of videos on how to author queries and fill in all the required/optional sections to achieve the query result as desired by the developer. Once the query is written, the developer can test the query by clicking on the run button. Figure 5c shows a typical interface for the query result with the main query and detail query results displayed one beneath the other. Once the developer is satisfied with the query results, the developer will choose to publish the query and direct the system administrator so that the administrator can make the query publicly available and would classify the query under the right topic in the QR Wiki table of contents.

## 5. Conclusions

In this paper, we present the QuranResearch.org system that targets a highly scalable cloud-based accurate database of the Quran that is accessible by all researchers over the world. We also explained the objectives and features of this system. The paper highlighted the system's interface from the perspective of the end user who can make simple inquiries or analyze the results of published queries, comment on them and start a healthy discussion. These discussions derive sound perceptions based on the computational aspect of the Quran research. On the other hand, the paper presented the system's interface from the developer's perspective, who is able to perform advanced queries in SQL language on the database, publish results that are accompanied by documentation. Developers and researchers are the users that the system relies on to add new modules and expand the system dynamically.

## Biodata

| | |
|---|---|
| 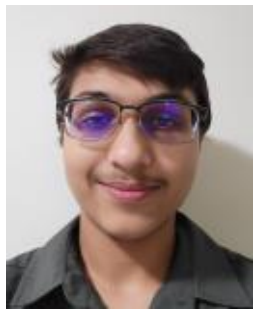 | Umar Ahmed Siddiqui is a student research assistant at the School of Engineering and Technology in University of Washington, Tacoma. Umar has built the developer's dashboard of Quranresearch.org website and provided training materials for the prospective developers of the system. Umar's research interests include AI, Spatio-temporal Databases, and Block-chain technology. |
| | Habiba Youssef is a student at the University of Washington double majoring in Psychology and Law, Economics & Public Policy. Her research interests include Muslim youth behaviour and understanding, Muslims' unique needs pertaining to mental health and the Islamic faith in hopes to connect Islamic teachings with therapy practices to support individuals on their mental, emotional and spiritual journeys. |
| 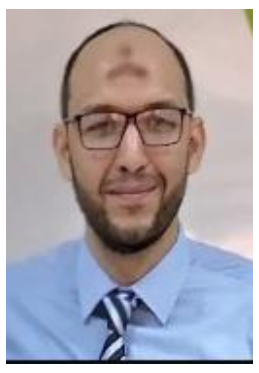 | Adel Sabour is a PhD student in Computer Science and Systems at the School of Engineering and Technology, University of Washington, Tacoma (UWT). Adel received his MS degree in Department of Computer Science and Information at the Institute of Statistical Studies and Research, Cairo University, Egypt. His research interests center around the fields of data mining, machine learning, NLP, computer vision, database/spatial database systems and big data. |
| 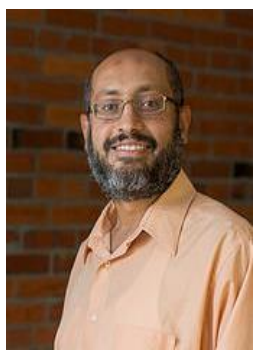 | Mohamed Ali is a professor of Computer Science and Systems at the School of Engineering and Technology, University of Washington, Tacoma (UWT). Mohamed received his PhD degree in computer science from Purdue University. Mohamed's research interests include Data Science, Big Data Systems, Database Management Systems and Spatiotemporal Databases. |

## Abstract in Arabic

# قابلية التوسع والتوافر وإعادة إنشاء النتائج والتمدد في أنظمة قواعد البيانات الإسلامية

**عمر صديقى[1]، حبيبة يوسف[2]، عادل صبور[3]، محمد علي[4]**

[1،2،3،4] قسم هندسة الحاسب والنظم، جامعة واشنطن، تاكوما، الولايات المتحدة الأمريكية

umarqr@uw.edu[1]

hyoussef@uw.edu[2]

sabour@uw.edu[3]

mhali@uw.edu[4]

**الخلاصة.** مع انتشار أنظمة وتطبيقات البرمجيات التي تخدم مجال المعرفة الإسلامية ، تبرز العديد من المخاوف. إن مصداقية ودقة قواعد البيانات التي تدعم هذه الأنظمة أمر يثير التساؤلات. مع الحماسة التي قد تكون لدى بعض مطوري البرامج والباحثين الهواة ، قد يتم تقديم بيانات كاذبة وادعاءات غير صحيحة حول الدلالات الرقمية أو المعجزات في القرآن. ولا تقدم أى طريقة لإعادة إنتاج هذه النتائج من قبل الأشخاص الذين يتبنون مثل هذه الادعاءات. علاوة على ذلك ، مع زيادة عدد المستخدمين ، أصبحت قابلية التوسع وتوافر هذه الأنظمة مصدر قلق. بالإضافة إلى كل هذه المخاوف ، فإن القابلية للتمدد هي أيضًا قضية رئيسية أخرى. يمكن أن تكون الأنظمة المصممة بشكل صحيح قابلة للتوسع وقابلة لإعادة الاستخدام ومبنية فوق بعضها البعض ، بدلاً من بناء كل نظام من نقطة الصفر في كل مرة يتم فيها بناء إطار عمل جديد. في هذا البحث ، نقدم لكم نظام QuranResearch.Org ورؤيته لقابلية التوسع والتوافر ، وإعادة إنشاء النتائج والتمدد لخدمة أنظمة قواعد البيانات الإسلامية.